\documentclass[preprint,english,letterpaper,showpacs]{revtex4}
\usepackage{babel,amsmath,amssymb}

\begin{document}

\title{Discrete Time Evolution and Energy Nonconservation in
  Noncommutative Physics}

\author{A. P. Balachandran}
\email{bal@phy.syr.edu}
\affiliation{Department of Physics, Syracuse University \\
Syracuse, NY, 13244-1130, USA}
\author{A. G. Martins}
\email{andrey@fis.unb.br}
\author{P. Teotonio-Sobrinho}
\email{teotonio@fma.if.usp.br}
\affiliation{Instituto de F\'{\i}sica, Universidade de S\~{a}o Paulo \\
C.P. 66318, S\~{a}o Paulo, SP, 05315-970, Brazil}

\pacs{03.65.Ca,11.10.Nx}

\preprint{SU-4252-844}

\begin{abstract}
Time-space noncommutativity leads to quantisation of time
\cite{Chaichian}  and energy nonconservation \cite{Andrey} when time
is conjugate to a compact spatial direction like a circle. 
In this context
energy is conserved only modulo some fixed unit.
Such a possibility arises for example in theories with a compact extra
dimension with which time does not commute. The above results
suggest striking phenomenological consequences in extra dimensional
theories and elsewhere. In this paper we develop scattering theory
for discrete time translations. It enables the calculation of
transition probabilities for energy nonconserving processes  and has a
central role both in formal theory and phenomenology.

We can also consider space-space noncommutativity where one of the
spatial directions is a circle. That leads to the quantisation of the
remaining spatial direction and conservation of momentum in that
direction only modulo some fixed unit, as a simple adaptation of the results
in this paper shows.

\end{abstract}

\maketitle

\section{Introduction}

In most approaches to quantum theory ${\it time}$ 
is a real parameter occurring in the evolution operator. On the
other hand, spatial coordinates can be seen as operators acting on a
Hilbert space of states.  

Many authors have recently considered quantum field theories defined
on noncommutative space-times. In this context time is
another coordinate, and together with the other coordinates it
generates a noncommutative $C^*$-algebra. Ideas coming from
noncommutative geometry have been extensively used in this research program.

In one of the most studied models, the noncommutative space-time
$\mathbb{R}^{D}_{\theta}=\mathcal{A}_{\theta}(\mathbb{R}^{D})$ is 
described by a deformation of the algebra of functions over 
$\mathbb{R}^{D}$, the deformed product being the Groenewold-Moyal (GM)
product between complex-valued functions. 
It has been claimed that due to the nonlocal character of the
GM product, QFT's constructed in this way have
intrinsic unitarity problems \cite{non_unitary}. Some authors have
tried to 
avoid this problem reformulating the rules leading to the scattering
amplitudes \cite{Bahns}. The authors of \cite{Doplicher} have shown
how to construct unitary quantum field theories using noncommutative
coordinates of the canonical (GM) type. Our approach to quantum
physics started in \cite{Molina,Andrey} is inspired by
\cite{Doplicher}.  

In \cite{Andrey}, we studied a class of noncommutative
space-times where time evolution is discrete.
The simplest example is the noncommutative cylinder
$\mathcal{A}_{\theta}(\mathbb{R}\times S^1)$.
We describe this space as the noncommutative unital $C^*$-algebra
generated by $\hat{x}_{0}$ and $e^{i\hat{x}_{1}}$,
with the defining relation:
\begin{equation}\label{basic_relation} 
e^{i\hat{x}_{1}} \hat{x}_{0}= 
\hat{x}_{0}e^{i\hat{x}_{1}} + \theta e^{i\hat{x}_{1}}\,.
\end{equation}
This can be seen as naturally coming from the canonical commutation relation 
$[\hat{x}_{0} , \hat{x}_{1}]=i\theta\mathbb{I}$ defining the GM
plane $\mathbb{R}^2_{\theta}=\mathcal{A}_{\theta}(\mathbb{R}^{2})$ 
when restricted to periodic functions of
$\hat{x}_{1}$.

The relation (\ref{basic_relation}) has two direct consequences: the
spectrum of the time coordinate $\hat{x}_{0}$ is discrete \cite{Chaichian},
being spaced by the fundamental time scale $\theta$, and more
interestingly, the time evolution is given by integral powers of 
the evolution operator corresponding to a time interval equal to $\theta$.  
This fact leads to the 
possibility of energy nonconservation in scattering processes
\cite{Andrey}: it is conserved only modulo  $\frac{2\pi}{\theta}$,
as we shall see in this paper. (See
also \cite{Bal_Chandar,Matschull,Hooft} in this connection.)

It would be natural to think of $\mathcal{A}_{\theta}(\mathbb{R}\times
S^1)$ as a two-dimensional noncommutative sub-space of a larger
space-time, the noncommuting coordinate being interpreted as a warped
space-like extra dimension. In this picture, discretisation of time
translation and attendant energy nonconservation would have
significant phenomenological consequences which wait to be explored.
This space-time picture is appropriate from
a field-theoretical point of view as well.
But in this paper we are concerned
about quantum mechanics, so that we can also interpret the
noncommuting variable as a sort of extra degree of freedom, which does
not commute with the time coordinate. 


Let $Q\times S^{1}$ be the configuration space of a quantum system, 
$Q$ being an ordinary $D$-dimensional configuration space.
Consider the algebra 
$\mathcal{A}$ generated by $\hat{q}_{1} \cdots , \hat{q}_{D}$ ,
$e^{i\hat{x}_{1}}$ and 
$\hat{x}_{0}$. We have:
\begin{equation}\label{algebra}
\mathcal{A}=\mathcal{F}(Q)
\otimes\mathcal{A}_{\theta}\left(\mathbb{R}\times S^1\right)
\,\,\,,\,\,\,\hat{q}_{i}e^{i\hat{x}_{1}}=e^{i\hat{x}_{1}}\hat{q}_{i}\,,
\end{equation}
where $\mathcal{F}(Q)$ is the commutative algebra of functions over
the configuration space $Q$, generated by the coordinates (coordinate
functions)     
$\,\hat{q}_{1} ,
\cdots , \hat{q}_{D}$. The only noncommutative piece of
$\mathcal{A}$ is the noncommutative cylinder. 

The algebra $\mathcal{A}$ is the noncommutative analogue of the space
of (time dependent) wave functions of ordinary quantum
theory. We point out that the operators
 $\,\hat{q}_{1}, \cdots , \hat{q}_{D}\,$ and $\,e^{i\hat{x}_{1}}\,$ play the
role of ``coordinates''. Just like in the usual case, we can define
position and momentum operators acting on elements of $\mathcal{A}$.  

The position operators associated to $Q$ are the same as in 
ordinary quantum theory. They are denoted by  
$\,\hat{q}^{L}_{1}, \cdots ,
\hat{q}^{L}_{D}$, the superscript ${}^L$ meaning left
multiplication. We can also introduce the multiplication
operator $e^{i\hat{x}_{1}^{L}}$.
Let $\hat{p}_{i}$ be the momentum canonically conjugate to
$\hat{q}^{L}_{i}$ and 
let $\hat{P}_{1}$ be the noncommutative analogue of $-i\partial_{1}$,
defined by
\begin{equation}
\hat{P}_{1}e^{i\hat{x}_{1}}=e^{i\hat{x}_{1}}\,.
\end{equation}
If $\hat{H}$ is the Hamiltonian operator describing a quantum system
based on $\mathcal{A}$, we can write it generically as:
\begin{equation}
\hat{H}=\hat{H}\left(\hat{q}^{L}_{1} ,\cdots , \hat{q}^{L}_{D}
,\hat{p}_{1} ,\cdots , \hat{p}_{D} , e^{i\hat{x}_{1}^L} , \hat{P}_{1}\right)\,.
\end{equation}

As we shall see later, it is the dependence of $\hat{H}$ on
$e^{i\hat{x}_{1}^L}$ and $\hat{P}_{1}$ that causes the discretization of time
translation and time evolution controlled by $\hat{H}$. 

A variant of the noncommutative  spacetime  model considered here
would be one where two spatial directions do not commute  and one of
them is a circle. Then the remaining spatial direction, which we can
identify with $\hat{x}_{0}$ in (\ref{basic_relation}), gets quantised
in units of $\theta$ and momentum in that direction is conserved only
modulo $\frac{2\pi}{\theta}$. The proofs of these results also follow from the
considerations of this paper.

\section{The Algebra}\label{section_algebra}

In this section we study the algebraic structure of $\mathcal{A}$. Its
noncommutative piece is responsible for all its interesting aspects,
so let us recall some facts about the noncommutative cylinder
$\mathcal{A}_{\theta}(\mathbb{R}\times S^1)$.
It follows from (\ref{basic_relation}) that
\begin{equation}
e^{i\frac{2\pi}{\theta}\hat{x}_{0}}e^{i\hat{x}_{1}}=
e^{i\hat{x}_{1}}e^{i\frac{2\pi}{\theta}\hat{x}_{0}}\,.
\end{equation}
Hence in every unitary irreducible representation of
$\mathcal{A}_{\theta}(\mathbb{R}\times S^1)$,
the central element $e^{i\frac{2\pi}{\theta}\hat{x}_{0}}$ is just a phase:
\begin{equation}\label{phase}
e^{i\frac{2\pi}{\theta}\hat{x}_{0}} =
e^{i\varphi}\mathbb{I} \,.
\end{equation}
For the spectrum $\textrm{spec} \, \hat{x}_{0}$ of
$\hat{x}_{0}$ in such a representation, we have from (\ref{phase}),
\begin{equation}
\textrm{spec} \, \hat{x}_{0} = 
\theta\left(\mathbb{Z}+\frac{\varphi}{2\pi}\right)
\equiv \left\{\theta
\left(n+\frac{\varphi}{2\pi}\right):n\in\mathbb{Z}\right\} \,, 
\end{equation}
which leads to the important relation:
\begin{equation}\label{important_relation}
e^{i\left(\omega+\frac{2\pi}{\theta}\right)\hat{x}_{0}}=
e^{i\varphi}\,e^{i\omega\hat{x}_{0}}\,.  
\end{equation}

We can realize
$\mathcal{A}_{\theta}(\mathbb{R}\times S^1)$ irreducibly
on an auxiliary Hilbert space $L^2(S^1)$. 
On this space, $e^{i\hat{x}_{1}}$ acts like a multiplication operator,
\begin{equation}
\left(e^{i\hat{x}_{1}}\alpha\right)\left(e^{ix_{1}}\right) =
e^{ix_{1}}\,\alpha\left(e^{ix_{1}}\right) \,,
\end{equation}
while $-\hat{x}_{0}/\theta$ acts like an ordinary momentum operator,

\begin{equation}
\left(\hat{x}_{0}\alpha\right)\left(e^{ix_{1}}\right)=
i\theta \frac{\partial \alpha}{\partial x_{1}}\left(e^{ix_{1}}\right)\,.
\end{equation}
We denote this representation as
$\mathcal{A}_{\theta}(\mathbb{R}\times S^1 , e^{i\frac{\varphi}{2\pi}}
)$. The flux term $\frac{\varphi}{2\pi}$ classifies all unitary
irreducible representation of
$\mathcal{A}_{\theta}(\mathbb{R}\times S^{1})$, every such
representation being equivalent to
$\mathcal{A}_{\theta}(\mathbb{R}\times S^1 , e^{i\frac{\varphi}{2\pi}} 
)$ for some value of $\varphi$. 

In what follows we deal directly with 
$\mathcal{A}_{\theta}(\mathbb{R}\times S^1 , e^{i\frac{\varphi}{2\pi}}
)$, instead of the algebra $\mathcal{A}_{\theta}(\mathbb{R}\times S^1
)$ itself. In order to do this,
we need an explicit description of
elements of $\mathcal{A}_{\theta}(\mathbb{R}\times S^1 ,
e^{i\frac{\varphi}{2\pi}})$. 
The first thing we note is that
only integral powers of $e^{i\hat{x}_{1}}$ belong to this
representation. In fact, it 
follows from (\ref{basic_relation}) that
%
%
%
%
%
\begin{equation}\label{integral_power}
e^{i\alpha\hat{x}_{1}}e^{i\lambda\hat{x}_{0}}e^{-i\alpha\hat{x}_{1}}=
e^{i\lambda(\hat{x}_{0}+\alpha\theta)}\,.
\end{equation}
Conjugating the relation (\ref{important_relation}) by
$e^{i\alpha\hat{x}_{1}}$ and using (\ref{phase}) and
(\ref{integral_power}), we get
\begin{equation}
e^{i2\pi\alpha}=1\,\,\,\,\mbox{or}\,\,\,\,\alpha\in\mathbb{Z}\,.
\end{equation}

Now consider an element of the subalgebra generated by
$\hat{x}_{0}$. We can write it generically as
\begin{equation}
\hat{\alpha}=\int_{-\infty}^{\infty}d\omega\,\tilde{\alpha}(\omega)
e^{i\omega\hat{x}_{0}}\,.
\end{equation}
Let us split this integral into an infinite sum of integrals, each
one taken over a finite interval of length $\frac{2\pi}{\theta}$,
\begin{eqnarray}\label{general_time}
\nonumber\hat{\alpha}=
\sum_{m\in\mathbb{Z}}\int_{(2m-1)\frac{\pi}{\theta}}
^{(2m+1)\frac{\pi}{\theta}}d\omega
\tilde{\alpha}\left(\omega\right)
e^{i\omega\hat{x}_{0}}&=&
\sum_{m\in\mathbb{Z}}\int_{-\frac{\pi}{\theta}}^{+\frac{\pi}{\theta}}d\omega
\tilde{\alpha}\left(\omega+m\frac{2\pi}{\theta}\right)
e^{i m \frac{2\pi}{\theta}\hat{x}_{0}}
e^{i\omega\hat{x}_{0}}=\\
&=&\int_{-\frac{\pi}{\theta}}^{+\frac{\pi}{\theta}}d\omega
\sum_{m\in\mathbb{Z}}\tilde{\alpha}
\left(\omega+m\frac{2\pi}{\theta}\right)e^{i m \varphi}\,
e^{i\omega\hat{x}_{0}}\,,
\end{eqnarray}
where we have made use of (\ref{phase}). Notice that the coefficient 
$$
f_{\tilde{\alpha}}\left(\omega\right):=\sum_{m\in\mathbb{Z}}\tilde{\alpha}
\left(\omega+m\frac{2\pi}{\theta}\right)e^{i m \varphi}
$$
in (\ref{general_time}) is quasi-periodic:
\begin{eqnarray}
f_{\tilde{\alpha}}\left(\omega+\frac{2\pi}{\theta}\right)=
e^{-i\varphi}f_{\tilde{\alpha}}(\omega)\,. 
\end{eqnarray}
From this we conclude that the most general element of
$\mathcal{A}_{\theta}(\mathbb{R}\times S^1 , e^{i\frac{\varphi}{2\pi}})$    
can be written as
\begin{equation}
\hat{\psi} =
\sum_{n\in\mathbb{Z}} 
\int_{-\frac{\pi}{\theta}}^{+\frac{\pi}{\theta}}\,d\omega\, 
\tilde{\psi}_{n}(\omega)\,e^{in\hat{x}_{1}}
e^{i\omega\hat{x}_{0}}\,,
\end{equation}
where 
$\tilde{\psi}_{n}(\omega+\frac{2\pi}{\theta})=
e^{-i\varphi}\tilde{\psi}_{n}(\omega)$. 

Taking (\ref{algebra}) into account we write the following expression
for the generic element 
of $\tilde{\mathcal{A}}:=\mathcal{F}(Q)
\otimes\mathcal{A}_{\theta}\left(\mathbb{R}\times S^1,
e^{i\frac{\varphi}{2\pi}}\right)$ \cite{comment}:  
\begin{equation}\label{general}
\hat{\psi} = 
\sum_{n\in\mathbb{Z}}  \int_{-\infty}^{+\infty} d^{D}k
\int_{-\frac{\pi}{\theta}}
^{+\frac{\pi}{\theta}}\,d\omega\,\tilde{\psi}_{n}(\vec{k} , \omega)
e^{i\sum_{i}k_{i}\hat{q}_{i}}\,e^{in\hat{x}_{1}}
e^{i\omega\hat{x}_{0}}\,,
\end{equation}
where $\tilde{\psi}_{n}(\vec{k} ,
\omega+\frac{2\pi}{\theta})=e^{-i\varphi}\tilde{\psi}_{n}(\vec{k} ,
\omega)$. 

\subsection{Translation Automorphisms}

In analogy with the common practice in quantum
mechanics, wave functions will be written in the ``basis'' 
$\{e^{i\sum_{i}k_{i}\hat{q}_{i}} , e^{in\hat{x}_{1}}\}$, just like in
(\ref{general}). In order to proceed with the construction of a
noncommutative quantum theory, we have to study the action of time and
space translations on $\tilde{\mathcal{A}}$. 
The product structure
(\ref{algebra}) shows that the momenta of $\mathcal{F}(Q)$
act on $\tilde{\mathcal{A}}$
as in the ordinary case. It remains to investigate the action of the
momenta $\hat{P}_{0}$ and $\hat{P}_{1}$, which act on
the GM plane
$\mathbb{R}^{2}_{\theta}=\mathcal{A}_{\theta}(\mathbb{R}^{2})$ as:
\begin{eqnarray}
\hat{P}_{0}=-\frac{1}{\theta}[\hat{x}_{1}\,,\cdot]\quad,\quad
\hat{P}_{1}=-\frac{1}{\theta}[\hat{x}_{0}\,,\cdot]\,,
\end{eqnarray}
so that
\begin{equation}\label{quantum_momenta}
\hat{P}_{\mu}\hat{x}_{\nu}=i\,\eta_{\mu \nu}\mathbb{I}\,,
\end{equation}
where $\eta_{\mu \nu}=\mbox{diag} \,[1 , -1]$.

Using the Jacobi identity (fulfilled by the commutator), one can show
that the momenta $\hat{P}_{\mu}$ generate a commutative
algebra. $\hat{P}_{\mu}$ are quite similar to their usual counterparts
$i\partial_{0}$ and $-i\partial_{1}$.

The only restriction on the action of $\hat{P}_{0}$ on $\tilde{\mathcal{A}}$
comes from the quasi-periodicity of the
coefficients $\tilde{\psi}_{n}(\vec{k} , \omega)$ in
(\ref{general}). In fact, let us calculate the action of an arbitrary
time translation on (\ref{general}):
\begin{eqnarray}
\nonumber e^{-i \tau \hat{P}_{0}}\hat{\psi}(M)&=&\sum_{n\in\mathbb{Z}}  
\int_{-\infty}^{+\infty} d^{D}k
\int_{-\frac{\pi}{\theta}}^{+\frac{\pi}{\theta}}\,d\omega\,
\tilde{\psi}_{n}(\vec{k} , \omega)
e^{i\sum_{i}k_{i}\hat{q}_{i}}\,e^{in\hat{x}_{1}}
e^{i\omega(\hat{x}_{0}+\tau\mathbb{I})}=\\
&=&\sum_{n\in\mathbb{Z}}  \int_{-\infty}^{+\infty} d^{D}k
\int_{-\frac{\pi}{\theta}}^{+\frac{\pi}{\theta}}\,d\omega\,
\tilde{\psi}_{n}(\vec{k} , \omega)e^{i\omega \tau}
e^{i\sum_{i}k_{i}\hat{q}_{i}}\,e^{in\hat{x}_{1}}
e^{i\omega\hat{x}_{0}}\,.
\end{eqnarray} 
If $e^{-i \tau \hat{P}_{0}}$ is to act
on $\tilde{\mathcal{A}}$, then the new coefficients $\tilde{\psi}_{n}(\vec{k},
\omega)e^{i\omega \tau}$ must also be quasi-periodic functions of
$\omega$. This condition is only fulfilled if $\tau\in
\theta\mathbb{Z}$. This shows that the time translations on
$\tilde{\mathcal{A}}$ are powers of a minimum translation
$e^{-i\theta\hat{P}_{0}}\hat{x}_{0}=\hat{x}_{0}+\theta\mathbb{I}$. 

The spatial translations (generated by $\hat{P}_{1}$)
act on $\tilde{\mathcal{A}}$ without any restriction.

\section{Dynamics and Hilbert Spaces}

In this section we recall the construction of Hilbert spaces in the
present approach to quantum physics.

\subsection{Symbols and Positive Functionals}

The first step is the introduction of 
a linear application from $\tilde{\mathcal{A}}$ to the space of complex
functions. We   
associate to every $\hat{\psi}$ as in (\ref{general}) the function
$\psi:\mathbb{R}^{D}\times
S^{1}\times\theta\left(\mathbb{Z}+\frac{\varphi}{2\pi}\right)\longrightarrow
\mathbb{C}$ given by:
\begin{eqnarray}\label{symbol}
\psi\left(\vec{q} , e^{ix_{1}} , 
\theta\left(m+\frac{\varphi}{2\pi}\right)\right) = 
\sum_{n\in\mathbb{Z}}  \int_{-\infty}^{+\infty} d^{D}k
\int_{-\frac{\pi}{\theta}}^{+\frac{\pi}{\theta}}\,
d\omega\,\tilde{\psi}_{n}(\vec{k} , \omega)
e^{i\sum_{i}k_{i} q_{i}}\,e^{in x_{1}}
e^{i\omega\theta\left(m+\frac{\varphi}{2\pi}\right)}\,.
\end{eqnarray} 
Equation (\ref{symbol}) defines the {\it symbol} $\psi$
of the operator $\hat{\psi}$.

Now we define a family of positive linear functionals on $\tilde{\mathcal{A}}$:

\begin{equation}\label{positive}
S_{\theta\left(m+\frac{\varphi}{2\pi}\right)}(\hat{\psi})=
\int_{-\infty}^{+\infty}d^{D}q\,
\int_{0}^{2\pi}dx_{1}\,\psi\left(\vec{q} , e^{ix_{1}} , 
\theta\left(m+\frac{\varphi}{2\pi}\right)\right)\,.
\end{equation}
In order to verify that (\ref{positive}) is a positive functional, one first
checks that
%
\begin{equation}\label{condition_positive}
S_{\theta\left(m+\frac{\varphi}{2\pi}\right)}(\hat{\psi}^{*}\hat{\phi})
=\int_{-\infty}^{+\infty}d^{D}q\,\int_{0}^{2\pi}dx_{1}
\overline{\psi\left(\vec{q} , e^{ix_{1}} , 
\theta\left(m+\frac{\varphi}{2\pi}\right)\right)}\,
\phi\left(\vec{q} , e^{ix_{1}} , 
\theta\left(m+\frac{\varphi}{2\pi}\right)\right)\,,
\end{equation}
which shows that 
$S_{\theta\left(m+\frac{\varphi}{2\pi}\right)}(\hat{\psi}^{*}\hat{\psi})
\geq 0$.

It is natural to try to define an inner product in the following way:
\begin{equation}\label{inner_product_tentative}
\left(\hat{\psi} , \hat{\phi}\right)_
{\theta\left(m+\frac{\varphi}{2\pi}\right)}
 = S_{\theta\left(m+\frac{\varphi}{2\pi}\right)}(\hat{\psi}^{*}\hat{\phi})\,.
\end{equation}
This family of sesquilinear forms have the right linearity
properties, but fails to
give $\tilde{\mathcal{A}}$ a Hilbert space structure. For any given
value of $m$
we can find nontrivial null vectors. In fact, let $\psi$ be the symbol
of $\hat{\psi}\in\tilde{\mathcal{A}}$. If $\psi\left(\vec{q} , e^{ix_{1}} , 
\theta\left(m+\frac{\varphi}{2\pi}\right)\right) = 0$ for some {\it
  fixed} $m\in\mathbb{Z}$ and for {\it all} $(\vec{q} ,
e^{ix_{1}})\in\mathbb{R}^{D}\times S^{1}$, then it follows from 
(\ref{condition_positive}) that:
\begin{equation}
S_{\theta\left(m+\frac{\varphi}{2\pi}\right)}(\hat{\psi}^{*}\hat{\psi})=0\,.
\end{equation}
But $\psi\left(\cdot \,,\, \cdot \,,\, 
\theta\left(n+\frac{\varphi}{2\pi}\right)\right)$ need not be zero for
$n\neq m$ and hence $\hat{\psi}$ need not be zero. Such nonzero
$\hat{\psi}$ are nontrivial null vectors for this form.

Besides this feature, if $n\neq m$ the
forms labeled by $\theta\left(n+\frac{\varphi}{2\pi}\right)$ and
$\theta\left(m+\frac{\varphi}{2\pi}\right)$
are not directly related to each other. This
time-dependence of the ``inner product'' is not a desirable feature in an
approach to quantum theory which is to reduce to ordinary
quantum mechanics in the appropriate commutative limit.

\subsection{The Hilbert Space of States}

The solution to the two problems mentioned in the last section
consists in finding a subspace
$\mathcal{H}_{\theta}$ of 
$\tilde{\mathcal{A}}$ such that: (i) the only null vector in
$\mathcal{H}_{\theta}$ is 
$0$ and (ii) the family of positive sesquilinear forms
(\ref{inner_product_tentative}) collapses to a unique true inner
product. 

Recall that in usual quantum mechanics, the Hilbert space of states
is the space of square integrable functions over the configuration
space 
$\,Q\,$ $(\mathcal{H}:=L^{2}(Q))$. Let $\psi\in\mathcal{H}$,
so that 
$\psi(q)\,\in\mathbb{C},\,\,$
for all $q\in Q\,.\,$ Time translation applied to $\psi$ is
equivalent to time evolution, and is given by the action of the
evolution operator $\,e^{-i\tau H}\,$ on $\psi$, where $H$ is the
Hamiltonian operator. As $H$ is self-adjoint, the evolution operator
is unitary, thereby defining an isometry in $\mathcal{H}=L
^{2}(Q)$. Hence we see that the evolved wave function
$\psi(\cdot\,,\,\tau)$ belongs to the same Hilbert space as $\psi(\cdot)$.

In the noncommutative case, time translation is given by the
action of the translation operator 
$U(N\theta):=e^{-iN\theta\hat{P}_{0}}$ on elements
of $\tilde{\mathcal{A}}$. One can check that
$e^{-i N \theta\hat{P}_{0}}$ is not a  
unitary operator on $\tilde{\mathcal{A}}$ (for the sesquilinear form
(\ref{inner_product_tentative})). In fact, there are pairs
$(\hat{\psi}\,,\,\hat{\phi})\in\tilde{\mathcal{A}}\times\tilde{\mathcal{A}}$
such that:  
\begin{equation}
\left(e^{-iN\theta \hat{P}_{0}}\hat{\psi}\,,\,\hat{\phi}
\right)_{\theta\left(m+\frac{\varphi}{2\pi}\right)}\neq
\left(\hat{\psi}\,,\,e^{-iN\theta \hat{P}_{0}}\hat{\phi}
\right)_{\theta\left(m+\frac{\varphi}{2\pi}\right)}\,.
\end{equation}
Hence time translation $e^{-iN\theta \hat{P}_{0}}$
cannot be used
to identify different metric spaces
$\left(\tilde{\mathcal{A}}\,,\,(\cdot\,,\,\cdot)
_{\left(m+\frac{\varphi}{2\pi}\right)}\right)$ labelled by $m$ unless there is a
suitable  constraint. Such a constraint is inspired by standard
quantum mechanics. We next explain it. It is a generalization of
Schr\"{o}dinger's equation. 

Consider a time-independent Hamiltonian $\hat{H}$ (this means that 
$[\hat{P}_{0} , \hat{H}]=0$), written generically as:
\begin{equation}\label{typical_hamiltonian}
\hat{H}=\hat{H}(\vec{q} , e^{i\hat{x}_{1}} , \vec{p} ,  
\hat{P}_{1})\,,
\end{equation}
where $\vec{q}=(\hat{q}^{L}_{1} , \cdots , \hat{q}^{L}_{D})$, 
$\,\vec{p}=(\hat{p}_{1} , \cdots , \hat{p}_{D})$ and
$\hat{P}_{1}$ is the momentum operator defined in 
(\ref{quantum_momenta}).

We suppose that (\ref{typical_hamiltonian}) is {\it hermitian} in
the following sense:
\begin{equation}
\left(\hat{\psi} , \hat{H}
\hat{\phi}\right)_{\theta\left(m+\frac{\varphi}{2\pi}\right)}=
\left(\hat{H} \hat{\psi} , \hat{\phi}\right)_
{\theta\left(m+\frac{\varphi}{2\pi}\right)}\,.
\end{equation}

Now in \cite{Molina}, to resolve a similar problem, 
the Hilbert spaces were defined as subspaces of the
full noncommutative algebra ($\mathcal{A}_{\theta}(\mathbb{R}^{D+1})$ in
that case) fulfilling an algebraic condition, the noncommutative
analogue of Schr\"{o}dinger's equation. Here we cannot do this,
because the generator of time translation $\hat{P}_{0}$ does not act
on $\tilde{\mathcal{A}}$. But we showed in section
\ref{section_algebra} that
integral powers of $e^{-i\theta\hat{P}_{0}}$ do act on $\tilde{\mathcal{A}}$,
so that we can impose the following substitute for the Schr\"{o}dinger's
constraint:
\begin{equation}\label{constraint}
e^{-i\theta  \hat{P}_{0}}\hat{\psi}=e^{-i\theta  \hat{H}}\hat{\psi}\,.
\end{equation}

Let $\mathcal{H}_{\theta}$ denote the subspace of
$\tilde{\mathcal{A}}$ consisting only of solutions of
(\ref{constraint}):
\begin{equation}\label{consists_of_solutions}
\mathcal{H}_{\theta}
:=\left\{\hat{\psi}\in\tilde{\mathcal{A}}\,:\,
e^{-i\theta  \hat{P}_{0}}\hat{\psi}=e^{-i\theta  \hat{H}}\hat{\psi}\right\}\,.
\end{equation}

The elements of $\mathcal{H}_{\theta}$ are of the form:
\begin{equation}
\hat{\psi}=e^{-i\hat{x}_{0}^R\hat{H}}\,
\hat{\chi}\left(\vec{q},e^{i\hat{x}_{1}}\right)\,,
\end{equation}
where $\hat{x}_{0}^R$ is multiplication by $\hat{x}_{0}$ from the right.

We show below that the space $\mathcal{H}_{\theta}$  just defined is a
true Hilbert space, with inner product given by
(\ref{inner_product_tentative}). 

Suppose $\hat{\psi}$ is a null vector of 
$(\cdot\,,\,\cdot)_{\theta\left(n+\frac{\varphi}{2\pi}\right)}$ for
some fixed $n$. This means that if $\psi$ denotes the symbol of
$\hat{\psi}$ then:
\begin{equation}
\psi\left(\vec{q} , e^{ix_{1}} , 
\theta\left(n+\frac{\varphi}{2\pi}\right)\right)=0\,
\end{equation}
for all $\vec{q}$ and all $e^{ix_{1}}$.

Let $\psi_{m}$ denote the symbol of 
$e^{-im\theta\hat{P}_{0}}\hat{\psi}.\,$ 
We have:
\begin{equation}\label{A}
\psi_{m}\left(\vec{q} , e^{ix_{1}} , 
\theta\left(n+\frac{\varphi}{2\pi}\right)\right)=
\psi\left(\vec{q} , e^{ix_{1}} , 
\theta\left(n+m+\frac{\varphi}{2\pi}\right)\right)
\end{equation}

Using the fact that $\,\hat{\psi}\in\mathcal{H}_{\theta}\,$ we can write:
\begin{equation}\label{B}
\psi_{m}=e^{-i\theta m H}\,\psi\,,
\end{equation}
where $H$ is the $\theta=0$ counterpart of $\hat{H}$: $H\psi$ is the
symbol of $\hat{H}\hat{\psi}$. We point out that 
the  $\theta=0$ operator $H$ acts only on the
spatial arguments ($\vec{q}$ and $e^{ix_{1}}$) of $\psi$ and is
time-independent. (This
follows from the fact that $\hat{H}$ does not depend on $\hat{P}_{0}$
and commutes with $\hat{P}_{0}$. 

Using (\ref{A}) and (\ref{B}) we get:
\begin{equation}
\psi\left(\vec{q} , e^{ix_{1}} , 
\theta\left(n+m+\frac{\varphi}{2\pi}\right)\right)=
e^{-i\theta m H}\,\psi\left(\vec{q} , e^{ix_{1}} , 
\theta\left(n+\frac{\varphi}{2\pi}\right)\right)\,.
\end{equation}
Recalling now that $\psi$ is a null vector of
$(\cdot\,,\,\cdot)_{\theta\left(n+\frac{\varphi}{2\pi}\right)}$, we find:
\begin{equation}
\psi\left(\vec{q} , e^{ix_{1}} , 
\theta\left(n+m+\frac{\varphi}{2\pi}\right)\right)=0
\end{equation}
for all $\vec{q}$ and $e^{ix_{1}}$. As $m$ is an arbitrary integer, we
see that $\psi$ is identically zero. Consequently the Fourier
coefficients $\tilde{\psi}_{n}(\vec{k} , \omega)$ (see equation
(\ref{symbol})) are also identically 
zero. Using (\ref{general})  we finally conclude that
$\hat{\psi}=0$. This shows that the only null vector of
(\ref{inner_product_tentative}) is the zero vector. This proves
that the subspace  $\mathcal{H}_{\theta}$ defined in
(\ref{consists_of_solutions}) is a 
Hilbert space.

If $\hat{\psi}$ and $\hat{\phi}$ fulfill (\ref{constraint}) we have:
\begin{eqnarray}
\nonumber\left(\hat{\psi} ,
\hat{\phi}\right)_{\theta\left(m+\frac{\varphi}{2\pi}\right)}&=&
\left( e^{-i\theta (m-n)\hat{P}_{0}}\hat{\psi} , e^{-i\theta
  (m-n)\hat{P}_{0}}\hat{\phi}\right)_
{\theta\left(n+\frac{\varphi}{2\pi}\right)}=\\
&=&\left( e^{-i\theta (m-n)\hat{H}}\hat{\psi} , e^{-i\theta
  (m-n)\hat{H}}\hat{\phi}\right)_
{\theta\left(n+\frac{\varphi}{2\pi}\right)}=
\left(\hat{\psi} , \hat{\phi}\right)_
{\theta\left(n+\frac{\varphi}{2\pi}\right)}\,,
\end{eqnarray}
which shows that in the subspace defined by (\ref{constraint}) the
family of
products (\ref{inner_product_tentative}) restricted to
$\mathcal{H}_{\theta}$  collapses to one unique inner 
product. We denote this
product as $\langle\,\,\cdot\,\,| \,\,\cdot\,\,\rangle$. 

Time translations in $\mathcal{H}_{\theta}$ are implemented by integral
powers of the unitary evolution operator $e^{-i\theta
  \hat{H}}$. The inner product of two elements of
$\mathcal{H}_{\theta}$ is independent of time, i.e.:
\begin{eqnarray}
\langle e^{-i\theta\hat{P}_{0}}\hat{\psi} \,|\,
e^{-i\theta\hat{P}_{0}}\hat{\phi}\rangle=
\langle e^{-i\theta\hat{H}}\hat{\psi} \,|\,
e^{-i\theta\hat{H}}\hat{\phi}\rangle=
\langle \hat{\psi} \,|\,
\hat{\phi}\rangle\,,
\end{eqnarray}
just like in ordinary quantum physics.

Our construction of the Hilbert space of states depends on the choice
of a suitable Hamiltonian operator.
We remark that 
in spite of the apparent multiplicity of Hilbert spaces implied by
the multiplicity of Hamiltonians, these spaces are in fact equivalent
to each other. This can be seen as follows. Let $\hat{H}_{1}$ and 
$\hat{H}_{2}$ be two
Hamiltonians fulfilling the necessary conditions for the applicability
of the construction described above. Let $\mathcal{H}_{\theta}^1$ and
$\mathcal{H}_{\theta}^2$ be the Hilbert spaces based on $\hat{H}_{1}$ and     
$\hat{H}_{2}$, respectively. The elements of $\mathcal{H}^{2}_{\theta}$ are related
to  the elements of  $\mathcal{H}^{1}_{\theta}$ throught the action of the unitary
operator 
\begin{eqnarray}
\nonumber
&&U_{21}:=e^{-i\hat{x}_{0}^R\hat{H}_{2}}\,e^{i\hat{x}_{0}^R\hat{H}_{1}}\\
&&U_{21}:\mathcal{H}_{\theta}^{1}\rightarrow \mathcal{H}_{\theta}^{2}\,,
\end{eqnarray}
while an observable $\hat{T}_{2}$ acting on $\mathcal{H}_{\theta}^2$ is
related to an observable  $\hat{T}_{1}$ acting on $\mathcal{H}_{\theta}^1$
by conjugation by $U_{21}$:
\begin{eqnarray}\label{conjugation}
\hat{T}_{2}=U_{21}\,\hat{T}_{1}\,U_{21}^{\dagger}\,.
\end{eqnarray}

\section{Scattering Theory}

In this section, we set $Q=\mathbb{R}^D$   and assume the following form
for the Hamiltonian acting on $\mathcal{H}_{\theta}\simeq 
L^2\left(S^1\times\mathbb{R}^D\right)$:
\begin{eqnarray}
\hat{H}=\frac{\hat{P}_{1}^2}{2M}+\frac{\vec{\hat{p}}^{\,2}}{2M}+
V(\,\vec{\hat{q}}\,)\equiv \hat{H}_{0}+V(\,\vec{\hat{q}}\,)\,.
\end{eqnarray}
Here $\hat{H}_{0}$ is the ``free'' Hamiltonian.

We include no interaction in the noncommutative direction. This is for
simplicity. There are still striking physical effects from time
discreteness. Further the formalism generalises to more general forms
of interaction.

We remark that even though the circle $S^1$ is compact, there is no
difficulty with cluster decomposition and defining asymptotic states
if, as we also assume, $V$ falls off sufficiently rapidly as 
$|\vec{\hat{q}}|\to\infty$.

Both free and full Hamiltonians $\hat{H}_{0}$ and $\hat{H}$ play a
fundamental role in scattering theory. To each one, we can associate a
Hilbert space, but, as we already discussed, they are equivalent to
each other. We choose to work uniformly  with the Hilbert space
$\mathcal{H}^{0}$ associated with the free Hamiltonian $\hat{H}_{0}$
in what follows. This implies in particular that $\hat{H}$ will be
substituted by its image $\hat{H}'$ under conjugation by the
appropriate unitary operator where (see equation (\ref{conjugation})):
\begin{eqnarray}
\hat{H}'=e^{-i\hat{x}_{0}^R\hat{H}_{0}}\,\hat{H}\,
e^{i\hat{x}_{0}^R\hat{H}_{0}}\,.
\end{eqnarray}
Note that $\hat{H}_{0}$ is invariant under this conjugation.

Standard scattering theory (cf. refs. \cite{scattering,collision}) can be adapted readily to the present
case. We shall give just the essential details to derive the equation
for the scattering matrix. The latter is exact and amenable to
approximate solutions by perturbation series or other methods.

Let $|in\,,\,\alpha\rangle$ and $|out\,,\,\alpha\rangle$  denote the
{\it in}- and {\it out}-states characterised by quantum numbers
$\alpha$. They evolve by $\hat{H}'$ and are prepared at time 0.

By definition, if $|\,in\,,\,\alpha\rangle$ is evolved back to infinite
past, it will (strongly) approach the free state with quantum numbers
$\alpha$ evolving by $\hat{H}_{0}$:
\begin{eqnarray}
\lim_{\buildrel{N\to-\infty}\over{N\in\mathbb{Z}}}
e^{-i\theta N\hat{H}'}|\,in\,,\,\alpha\rangle=
\lim_{\buildrel{N\to-\infty}\over{N\in\mathbb{Z}}}
e^{-i\theta N\hat{H}_{0}}|\alpha\rangle\,.
\end{eqnarray}
Thus
\begin{eqnarray}
|in\,,\,\alpha\rangle=\lim_{\buildrel{N\to-\infty}\over{N\in\mathbb{Z}}}
\left(e^{i\theta N\hat{H}'}\,e^{-i\theta
  N\hat{H}_{0}}\right)|\alpha\rangle
=:\Omega^{(+)}|\alpha\rangle\,,
\end{eqnarray}
where
\begin{eqnarray}
\Omega^{(+)}=\lim_{\buildrel{N\to-\infty}\over{N\in\mathbb{Z}}}
\left(e^{i\theta N\hat{H}'}\,e^{-i\theta
  N\hat{H}_{0}}\right)
\end{eqnarray}
is the M\"{o}ller operator transforming the free to the {\it in} state.

Similarly by definition, $|\,out\,,\,\alpha\rangle$ if evolved to
infinite future will approach the free state with quantum numbers
$\alpha$ evolving by $\hat{H}_{0}$:
\begin{eqnarray}
\lim_{\buildrel{N\to+\infty}\over{N\in\mathbb{Z}}}
e^{-i\theta N\hat{H}'}|\,out\,,\,\alpha\rangle=
\lim_{\buildrel{N\to+\infty}\over{N\in\mathbb{Z}}}
e^{-i\theta N\hat{H}_{0}}|\alpha\rangle
\end{eqnarray}
or
\begin{eqnarray}
\nonumber|out\,,\,\alpha\rangle&=&\Omega^{(-)}|\alpha\rangle\,,\\
\Omega^{(-)}&=&\lim_{\buildrel{N\to+\infty}\over{N\in\mathbb{Z}}}
e^{i\theta N\hat{H}'}\,e^{-i\theta N\hat{H}_{0}}\,.
\end{eqnarray}

Certain essential features of $\Omega^{(\pm)}$ follow readily from their
definitions. Thus, 
\begin{eqnarray}
\nonumber e^{\pm i\theta\hat{H}'}\,\Omega^{(+)}&=&
\lim_{\buildrel{N\to-\infty}\over{N\in\mathbb{Z}}}
e^{i\theta (N\pm1)\hat{H}'}\,e^{-i\theta N\hat{H}_{0}}=\\
&=&\lim_{\buildrel{N'\to-\infty}\over{N\in\mathbb{Z}}}
e^{i\theta N'\hat{H}'}\,e^{-i\theta (N'\mp1)\hat{H}_{0}}
\end{eqnarray}
or
\begin{eqnarray}
e^{\pm i\theta\hat{H}'}\,\Omega^{(+)}=\Omega^{(+)}\,
e^{\pm i\theta\hat{H}_{0}}\,,
\end{eqnarray}
and similarly,
\begin{eqnarray}
e^{\pm i\theta\hat{H}'}\,\Omega^{(-)}=\Omega^{(-)}\,
e^{\pm i\theta\hat{H}_{0}}\,.
\end{eqnarray}

The $S$-matrix element for scattering from an initial to a final state
with quantum numbers $\alpha$ and $\beta$ respectively is
\begin{eqnarray}
S_{\beta \alpha}=\langle\,out\,,\beta\,|\,in\,,\,\alpha\,\rangle
=\langle\beta\,|\,{\Omega^{(-)}}^{\dagger}\Omega^{(+)}\,
|\,\alpha\,\rangle\,,
\end{eqnarray}
where
\begin{eqnarray}
S={\Omega^{(-)}}^{\dagger}\Omega^{(+)}=
\lim_{\buildrel{M\to +\infty}\over{M\in\mathbb{Z}}}
U_{I}(\theta\,,\,M\,,-\infty)\,,
\end{eqnarray}
\begin{eqnarray}\label{interaction}
U_{I}(\theta\,,\,M\,,-\infty)&=&e^{i\theta M\hat{H}_{0}}
e^{-i\theta M\hat{H}'}\,\Omega^{(+)}\,.
\end{eqnarray}
The operator $S$  is the interaction representation $S$-matrix.

We next extract the $\delta$-function in $S_{\beta \alpha}$
corresponding to energy conservation $mod\,\,\frac{2\pi}{\theta}\,$ as
follows. From  (\ref{interaction}),
\begin{eqnarray}\label{summing}
\nonumber
U_{I}(\theta\,,\,M+1\,,-\infty)&-&U_{I}(\theta\,,\,M\,,-\infty)=
e^{i\theta M\hat{H}_{0}}\,W\,\Omega^{(+)}\,e^{-i\theta M\hat{H}_{0}}\,,\\
&&W=e^{i\theta\hat{H}_{0}}\,e^{-i\theta\hat{H}'}-\mathbb{I}\,.
\end{eqnarray}
Summing both sides from $M=-L$ to $M=J$,
\begin{eqnarray}
U_{I}(\theta\,,\,J+1\,,-\infty)-U_{I}(\theta\,,\,-L\,,-\infty)=
\sum_{\buildrel{k=-L,}\over{k\in\mathbb{Z}}}^{J}e^{i\theta k\hat{H}_{0}}\,W\,\Omega^{(+)}\,
\,e^{-i\theta k\hat{H}_{0}}\,.
\end{eqnarray}
Using the boundary conditions
\begin{eqnarray}\label{boundary}
\lim_{\buildrel{L\to +\infty}\over{L\in\mathbb{Z}}}
U_{I}(\theta\,,\,-L\,,-\infty)=\mathbb{I}\,,
\end{eqnarray}
\begin{eqnarray}
\lim_{\buildrel{J\to +\infty}\over{J\in\mathbb{Z}}}
U_{I}(\theta\,,\,J+1\,,-\infty)=S\,,
\end{eqnarray}
we get
\begin{eqnarray}
S=\mathbb{I}+\sum_{\buildrel{k=-\infty,}\over{k\in\mathbb{Z}}}^{+\infty}
e^{i\theta k\hat{H}_{0}}\,W\,\Omega^{(+)}\,
\,e^{-i\theta k\hat{H}_{0}}\,.
\end{eqnarray}

Let $\alpha\,(\beta)$ denote energy
$E_{\alpha}\,\left(E_{\beta}\right)$ and possible additional quantum
numbers $\alpha'\,(\beta')$. We write $\alpha=E_{\alpha},\alpha'\,\,$,
$\,\beta=E_{\beta},\beta' $. Then
\begin{eqnarray}
S_{\beta \alpha}=\delta_{\beta \alpha}-i\theta\left(
\sum_{\buildrel{k=-\infty,}\over{k\in\mathbb{Z}}}^{k=+\infty}e^{i\theta k(E_{\beta}-E_{\alpha})}\right)
\,T_{\beta \alpha}\,,
\end{eqnarray}
\begin{eqnarray}
T_{\beta \alpha}=\frac{i}{\theta}\langle\, E_{\beta},\beta'\,|\,W
\,\Omega^{(+)}\,|\,E_{\alpha},\alpha'\rangle\,.
\end{eqnarray}

Making use of the identity
\begin{eqnarray}
\frac{2\pi}{\theta}\delta_{S^1}(x)=\sum_{n\in\mathbb{Z}}
e^{in\theta x}\,,
\end{eqnarray}
where $\delta_{S^1}$ is the periodic $\delta$-function with period
$\frac{2\pi}{\theta}$, we get the final form of $S_{\beta \alpha}$:
\begin{eqnarray}\label{delta_appears}
S_{\beta \alpha}=\delta_{\beta \alpha}-2\pi i\,\delta_{S^1}
(E_{\beta}-E_{\alpha})\,T_{\beta \alpha}\,.
\end{eqnarray}
{\it Since $T_{\beta \alpha}$ need not be zero if 
$E_{\beta}\neq E_{\alpha}$, this equation clearly shows that
  scattering conserves energy only $mod\,\,\frac{2\pi}{\theta}$.}

As $\,\theta\to 0$, $\,W\to
-i\theta\,(\hat{H}'-\hat{H}_{0})\,$ and
$\,T_{\beta \alpha}\,$ approaches the familiar form 
$\,\langle\, E_{\beta},\beta'\,|\,(\hat{H}'-\hat{H}_{0})
\,\Omega^{(+)}\,|\,E_{\alpha},\alpha'\rangle$. In this limit,
$\frac{2\pi}{\theta}\to \infty$. Since 
$\,(\hat{H}'-\hat{H}_{0})\,\Omega^{(+)}\,$ cannot cause transitions
which change energy by an infinite amount,
this matrix element
approaches zero as $\theta\to 0$ if $E_{\beta}$  differs
from $E_{\alpha}$ by a non-zero multiple of $\frac{2\pi}{\theta}$. Thus we may
put $E_{\beta}=E_{\alpha}$ as well as let $\,\theta\to 0$
thereby recovering the usual $\theta=0$ expression for
$T_{\beta \alpha}$. 

For cross-section calculations, the relevant operator is the
restriction of $\,\frac{i}{\theta}\,W\,\Omega^{(+)}\,$ to eigenstates of
$\hat{H}_{0}$ for energy eigenvalue $E_{\alpha}$. Its overlap with $\,|E_{\beta},\beta'\rangle\,$
then gives the transition amplitude $\,T_{\beta \alpha}\,$. Call this
restriction to energy $E$ states of $\hat{H}_{0}$ as $T(E)$:
\begin{eqnarray}
\nonumber
&&T(E):=\frac{i}{\theta}\,W\,\Omega^{(+)}\,P(E)\,,\\
P(E)&=&\mbox{projector to energy $E$ states}\,.
\end{eqnarray}

For $\theta\to 0$, $T(E)$ fulfills the Lippman-Schwinger
equation. We now derive its analogue for $\theta\neq 0$.

From (\ref{interaction}), we see that
\begin{eqnarray}\label{analogue}
U_{I}(\theta\,,\,0\,,\,-\infty)=\Omega^{(+)}\,.
\end{eqnarray}

Hence summing (\ref{summing}) from $-\infty$ to $-1$ and using
(\ref{boundary}) and (\ref{analogue}), we get
\begin{eqnarray}
\Omega^{(+)}(E+i\epsilon):=\Omega^{(+)}\,P(E)=\left(
\mathbb{I}+\sum_{\buildrel{k=-\infty,}\over{k\in\mathbb{Z}}}^{-1}
e^{i\theta k(\hat{H}_{0}-E-i\epsilon)}\,W\,\Omega^{(+)}\right)\,P(E)\,,
\end{eqnarray}
where we have put a small positive imaginary part $\epsilon$ in $E$ to
ensure convergence of the series. (This is standard procedure also for
$\theta=0$. The limit $\epsilon\to 0$ is finally taken.) Summing the
series, 
\begin{eqnarray}\label{summing_series}
\Omega^{(+)}(E+i\epsilon)=
\left[\mathbb{I}+\frac{1}{e^{i\theta(\hat{H}_{0}-E-i\epsilon)}-\mathbb{I}}\,
W\,\Omega^{(+)}(E+i\epsilon)\right]\,P(E)\,.
\end{eqnarray}

The generalised Lippman-Schwinger equation follows on multiplying
(\ref{summing_series}) by $\frac{i}{\theta}\,W$:
\begin{eqnarray}
T(E+i\epsilon)=\left[\frac{i}{\theta}\,W+\,W
\frac{1}{e^{i\theta(\hat{H}_{0}-E-i\epsilon)}-\mathbb{I}}\,T(E+i\epsilon)
\right]\,P(E)\,.
\end{eqnarray}

Let 
\begin{eqnarray}
W(E)=W\,P(E)\,.
\end{eqnarray}
Then this equation is
\begin{eqnarray}\label{L_S}
T(E+i\epsilon)=\frac{i}{\theta}W(E)+W\,
\frac{1}{e^{i\theta(\hat{H}_{0}-E-i\epsilon)}-\mathbb{I}}\,T(E+i\epsilon)\,.
\end{eqnarray}

Equation (\ref{L_S}) is well-adapted to expansion of $T(E+i\epsilon)$
in a power series in $W$. As $W$ is invariant under the substitutions 
$\,\hat{H}_{0}\to\hat{H}_{0}+\frac{2\pi}{\theta}\,n\,$, 
$\,\hat{H}'\to\hat{H}'+\frac{2\pi}{\theta}\,m\,$,
$n , m\,\in\mathbb{Z}\,$, an approximation based on this series is
compatible with energy conservation $\,mod\,\,\frac{2\pi}{\theta}$.

Let 
\begin{equation}
G^{(+)}(E+i\epsilon)=\frac{\theta}{i}\frac{1}{e^{i\theta(\hat{H}_{0}-E-i\epsilon)}-\mathbb{I}}\,.
\end{equation}

As $\theta\to 0$,
\begin{eqnarray}
G^{(+)}(E+i\epsilon)\to
\frac{1}{E+i\epsilon-\hat{H}_{0}}\,,
\end{eqnarray}
which is the time-independent Green's function of $\hat{H}_{0}$. Hence 
$G^{(+)}(E+i\epsilon)$ is the discrete analogue of this Green's
function.

As in standard scattering theory, we can also formally solve
(\ref{L_S}) for $T(E+i\epsilon)$:
\begin{eqnarray}
T(E+i\epsilon)=\frac{1}{\mathbb{I}-\left(\frac{i}{\theta}W\right)\,G^{(+)}(E+i\epsilon)}\,
\left(\frac{i}{\theta}W(E)\right)\,.
\end{eqnarray}
The expansion of the inverse in powers of $W$ gives the perturbation
series for $T(E+i\epsilon)$ in $W$.

The above scattering theory can be applied to a variety of
interesting problems. Just as an example, the Born approximation
$T_{BA}(E+i\epsilon)$ for the Yukawa potential
\begin{equation}\label{Yukawa}
V(r)=V_{0}\,\frac{e^{-r/a}}{r}
\end{equation}
reads
\begin{equation}
T_{BA}(E+i\epsilon)=\frac{i}{\theta}\left(e^{i\theta\hat{H}_{0}}
e^{-i\theta\left(\hat{H}_{0}+V_{0}\,\frac{e^{-r/a}}{r}\right)}-\mathbb{I}\right)
\,P(E)\,.
\end{equation}

The matrix element of $T_{BA}(E_{\alpha}+i\epsilon)$  between an initial free
state with energy $E_{\alpha}$ and a final state with energy $E_{\beta}\neq E_{\alpha}$
gives the transition matrix $T_{\beta\alpha}$. As long as  the Yukawa
potential does not commute with the free Hamiltonian $\hat{H}_{0}$,
$T_{\beta\alpha}$ is nontrivial, the Yukawa
interaction being responsible for transitions between states with different energies. The
observable transitions are the scattering channels allowed by the periodic
delta function $\delta_{S^1}
(E_{\beta}-E_{\alpha})$ in (\ref{delta_appears}). 

It should be noted that just as in the ordinary case, the parameter $V_{0}$ in (\ref{Yukawa})
controls the size of the perturbation $\frac{i}{\theta}W$.

\section{Conclusions}

The present paper shows explicitly that scattering processes with discrete time
evolution lead to energy
nonconservation. Specifically, energy is conserved only {\it mod}
$\frac{2\pi}{\theta}$, $\theta$ being the elementary time interval.

The time independent scattering theory developed in section IV allows
perturbation calculations, just like in ordinary quantum
theory. This fact suggests a variety of interesting applications of
this formalism.

In addition to the perturbative approach, it will be interesting to
construct exactly soluble models in the future, in order to give us
detailed insights on
the general behaviour of the transition matrix.

The results of this paper have the merit of permiting straightforward
calculations of transition rates. This fact will certainly lead to striking
phenomenological consequences which can be confronted with experiments.

\begin{acknowledgments}
This work was  supported by DOE under contract number DE-FG02-85ER40231,
by NSF under contract number INT9908763 and by FAPESP, Brazil. 
\end{acknowledgments}



\begin{thebibliography}{10}
\bibitem{comment} Notice that $\tilde{A}$ comes from
  $\mathcal{A}$ when we replace $\mathcal{A}_{\theta}(\mathbb{R}\times
  S^{1})$  by $\mathcal{A}_{\theta}(\mathbb{R}\times S^1 ,
  e^{i\frac{\varphi}{2\pi}})$.
\end{thebibliography}
\end{document}